\documentstyle[twoside,fleqn,espcrc2]{article}

\begin{document}

\title{Neutrino Gravitational Redshift and the Electron Fraction  
Above Nascent
Neutron Stars}
\author{George M. Fuller\address{Department of Physics, University of
California, San Diego,
La Jolla, California 92093-0319}
and Yong-Zhong Qian\address{Physics Department, 161-33,
California Institute of Technology, Pasadena, California 91125}}
\begin{abstract}
Neutrinos emitted from near the surface of the hot proto-neutron star
produced by a supernova explosion may be subject to significant
gravitational redshift at late times. Electron
antineutrinos (${\bar{\nu}}_e$) decouple deeper in the gravitational  
potential
well of the
neutron star than do the electron neutrinos (${\nu}_e$), so that the
${\bar{\nu}}_e$
experience a larger
redshift effect than do the ${\nu}_e$. We show how this differential
redshift can increase the electron fraction $Y_e$
in the neutrino-heated ejecta from the neutron star. Any $r$-process
nucleosynthesis
originating in the neutrino-heated ejecta would require a low $Y_e$,  
implying
that the differential
redshift effect cannot be too large. In turn, this effect may allow
nucleosynthesis to probe
the nuclear equation of state parameters which set the neutron star  
radius and
surface density scale height at times of order $t_{\rm pb} \approx  
10\,{\rm s}$
to $25\,{\rm s}$ after
core bounce.
\end{abstract}

\maketitle

In this paper we examine the effects of gravitational redshift on the  
energy
spectra of
neutrinos and antineutrinos emitted from the surface regions of the  
hot neutron
star remnants
of Type II and Type Ib supernova explosions. These considerations  
allow us to
find a serendipitous
link between the neutron-to-proton ratio required for $r$-process
nucleosynthesis in the
neutrino-heated supernova ejecta and the size and structure of the  
neutron star
at times of order
one or several neutrino diffusion timescales after core bounce.

This link is significant, because it may be
in just this time frame ($t_{\rm pb} \approx 10\,{\rm s}$ to  
$25\,{\rm s}$,
where $t_{\rm pb}$ stands for time
{\it post core bounce})
that enough deleptonization of the proto-neutron star
has occurred to trigger an exotic phase
transition to quark (strange) matter or, perhaps more likely, a kaon  
condensate
\cite{brown92}.
Kaon condensation, for example, would have a fairly dramatic effect  
on the
evolution of the deleptonizing neutron
star. As Gerry Brown and his co-workers have elucidated \cite{bb94},  
such
a phase transition almost inevitably leads to the
formation of a black hole.
Whether or not there is
an exotic phase transition, as the neutron star deleptonizes
its radius would shrink and we might expect general relativistic  
effects near
the star's
surface to become more pronounced.

Interestingly, calculations indicate that the neutrino-heated ejecta
originating
from the vicinity of the neutron star surface
at these late times is a leading candidate for the site of
$r$-process nucleosynthesis
\cite{meyer92,woosley94}.
In these calculations neutrinos not only
provide the requisite energy to eject this material
\cite{qianwoos96}, but the competition \cite{qian93} between electron  
neutrino
and antineutrino captures on neutrons and protons, respectively,
$${\nu}_e + {\rm n} \rightarrow {\rm p} + e^-,\eqno(1a)$$
$${\bar{\nu}}_e + {\rm p} \rightarrow {\rm n} +e^+ ,\eqno(1b)$$
sets the neutron-to-proton ratio n/p$=1/Y_e -1$, where $Y_e$ is
the electron fraction, or the net number of electrons per baryon.  
Integration
of the rate equations \cite{qian93} corresponding to the reactions in  
equations
$(1a)$ and $(1b)$
shows that,
$$Y_e \approx {\left[{1 + {\lambda}_{{\bar{\nu}}_e  
p}(r)/{\lambda}_{{\nu}_e
n}(r)}\right]}^{-1},\eqno(2a)$$
where ${\lambda}_{{\bar{\nu}}_e p}(r)$ and ${\lambda}_{{\nu}_e n}(r)$
correspond to the rates of the
reactions in equations $(1b)$ and $(1a)$, respectively, as evaluated  
at radius
$r$.  We can approximate this expression as,
$$Y_e \approx \left[1 + \left(L_{\bar\nu_e}
\langle E_{\bar\nu_e}\rangle\right)/\left(
L_{\nu_e}  
\langle E_{\nu_e}\rangle\right)\right]^{-1},\eqno(2b)$$
where $\langle{E_{{\bar{\nu}}_e}}
\rangle $ and $\langle{E_{{\nu}_e}}\rangle $ represent the average
${\bar{\nu}}_e$ and
${\nu}_e$ energies, respectively, while ${L_{{\bar{\nu}}_e}}$ and
${L_{{\nu}_e}}$ are the ${\bar{\nu}}_e$ and ${\nu}_e$ luminosities,
respectively. Equation $(2b)$ follows from equation $(2a)$ on noting  
that, for
example,
the cross section for the process in equation $(1a)$ scales like the  
square of
the neutrino energy, and that the rate for this process
is given by the neutrino number flux times the appropriate cross  
section with,
in turn, the number flux given by the ratio of the neutrino  
luminosity and the
appropriate neutrino average energy.

In the absence of neutrino flavor transformations and extreme  
gravitational
redshift effects, the average local energies of the various neutrino  
species
above the neutrino sphere obey the hierarchy:
$$\langle E_{{\nu}_{\mu(\tau)}}\rangle\approx \langle
E_{\bar\nu_{\mu(\tau)}}\rangle > \langle E_{{\bar{\nu}}_e}\rangle
> \langle E_{{\nu}_e}\rangle .\eqno(3)$$
This energy hierarchy reflects the fact that the neutron star is  
mostly
neutrons: the ${\nu}_e$ have a higher opacity contribution from the  
charged
current process in equation $(1a)$ than the ${\bar{\nu}}_e$ acquire  
from the
process in equation $(1b)$. As a result, the ${\bar{\nu}}_e$ tend to  
decouple
deeper in the star where it is hotter, and the ${\nu}_e$ tend to  
decouple
further out where it is relatively cooler. Since the muon and tau  
neutrinos and
antineutrinos lack any charged current opacity contribution, they  
decouple
deepest and thus have larger average energies than either of the  
electron-type
neutrino species. Typical average energies at $t_{\rm pb} \sim  
10\,{\rm s}$ are
\cite{qian93}: $\langle E_{{\nu}_{\mu(\tau)}}\rangle\approx\langle
E_{\bar\nu_{\mu(\tau)}}\rangle \approx 25\,{\rm MeV}$; $\langle
E_{{\bar{\nu}}_e}\rangle \approx 16\,{\rm MeV}$;
and $\langle E_{{\nu}_e}\rangle \approx 11\,{\rm MeV}$. At these  
times,
numerical calculations suggest that the luminosities of all six  
neutrino
species are nearly the same, so that by equation $(2b)$ it is clear  
that $Y_e <
0.5$ and neutron-rich conditions will obtain so long as the average  
energy
hierarchy is as in equation $(3)$.

Neutron-rich conditions, $Y_e < 0.5$, are {\it necessary} for  
$r$-process
nucleosynthesis to take place in the neutrino-heated ejecta generated  
in the
post-core-bounce supernova environment. However, current models of  
the
$r$-process based on this environment, though promising, are also  
problematic.
The $Y_e$,
expansion timescale, and entropy conditions predicted in these models  
are at
best barely adequate to give the neutron-to-seed  ratio required for  
the
production of nuclei like platinum.
It is very clear that these models would not be viable if $Y_e$ were  
to be {\it
increased} significantly over the predictions of equations $(2a)$ and  
$(2b)$
with the average neutrino energies as above. Though the neutrino  
energy spectra
are themselves rather uncertain as a result of, for example,  
transport
calculation deficiencies, one might hope that the fractional {\it  
difference}
between $\langle E_{{\bar{\nu}}_e}\rangle $ and $\langle  
E_{{\nu}_e}\rangle $,
which has the most leverage on $Y_e$, could be obtained somewhat more  
reliably.
Numerical models show that the fractional difference between $\langle
E_{{\bar{\nu}}_e}\rangle $ and $\langle E_{{\nu}_e}\rangle $  
increases with
time, as would be expected with continuing deleptonization and the  
concomitant
lessening of the charged current opacity contribution [equation  
$(1a)$] for the
${\bar{\nu}}_e$. It might be tempting to argue that the $r$-process  
in the
neutrino-heated ejecta comes from very late times post-core-bounce,  
when this
electron antineutrino/neutrino fractional average energy difference  
is large
and, therefore, $Y_e$ in the ejecta is small.

But this may be dangerous, as invoking very late times and  
significant
deleptonization raises the possibility that gravitational redshift  
effects can
become important, especially if there is a phase transition to an  
exotic
equation of state.
At this epoch we expect the radius of the neutron star to change only  
slowly
with time, essentially on a neutrino diffusion timescale. Therefore,  
to gauge
the effects of redshift on the neutrino and antineutrino energy  
spectra it is
adequate to approximate the spacetime as static. We will also neglect  
rotation
and magnetic fields, and assume that spherical symmetry is a good
approximation. With these approximations, we can describe adequately  
the region
outside the neutron star as the vacuum Schwarzschild geometry with  
metric
components $g_{{\alpha} {\beta}}$ defined through the line element,  
${d s}^2 =
g_{{\alpha} {\beta}} {d x^{\alpha}} {d x^{\beta}}$,
$${d s}^2 = -e^{\left\{{2\Phi (r)}\right\}} {d t}^2 +  
e^{\left\{{2\Lambda
(r)}\right\}} {d r}^2 +
r^2 {d \Omega}^2,\eqno(4)$$
where ${d \Omega}^2  = ( {d \theta}^2 +\sin^2\theta {d \phi}^2 )$,  
and $\left\{
x^{\alpha} \right\} \rightarrow$ ($t$, $r$, $\theta$, $\phi$) are the
Schwarzschild coordinates. Here $\Phi (r)$ and $\Lambda (r)$ are the  
usual
radial coordinate-dependent Schwarzschild metric functions  
\cite{mtw}. In
vacuum the metric function in the time-time component of the metric  
is given by
${\exp}\left\{{2\Phi (r)}\right\} = 1 - r_s/r$, where the  
Schwarzschild radius
is $r_s = 2M \approx 4.134\,{\rm  
km}\left({{M}/{1.4\,{M_{\odot}}}}\right)$,
with $M$ the gravitational mass of the neutron star, and where we  
have set
$G=c=1$.

In any static spacetime, the timelike {\it covariant} component of  
the
four-momemtum $p_0 = g_{0\beta} p^{\beta}$ [where ${\bf p}  
\Rightarrow \left\{
p^\beta\right\} \rightarrow $ ($p^0$, $p^1$, $p^2$, $p^3$)] of a  
freely falling
particle is a constant of the motion for covering coordinates in  
which the
metric functions are time independent, {\it e.g.} the Schwarzschild
coordinates. Neutrinos emitted from a \lq\lq neutrino sphere\rq\rq\  
near the
neutron star's surface can be taken to be freely streaming, and thus  
following
geodesics, in the region well above the surface. The energy $E^{\ast}  
(r)$
measured by a locally inertial observer (with four velocity ${\bf  
u}$) at some
point at radius $r$ along a neutrino trajectory will be $E^{\ast} (r)  
= -{\bf
p}\cdot {\bf u}$.

If we take this observer to be momentarily at rest at position $r$,  
and
evaluate the frame invariant inner product in the Schwarzschild  
coordinates [so
that the four velocity has components ${\bf u} \rightarrow$
(${\exp}\left\{{-\Phi}\right\}$, $0$, $0$, $0$)], then the locally  
measured
energy will be,
$$E^{\ast} (r) = E {\exp}\left\{{-\Phi (r)}\right\} ,\eqno(5a)$$
where $E$ is a conserved quantity defined in terms of the time-like  
covariant
momentum component in the Schwarzschild coordinates as $E \equiv  
-{p_0}$. With
this definition, $E$ is interpreted as the energy of the neutrino as  
measured
at infinity.
In terms of the neutrino production (emission) energy at the neutrino  
sphere,
$E_{\rm em}$, as measured by a locally inertial observer at rest, we  
can write,
$$E = {E_{\rm em}} {\exp}\left\{{\Phi (r_{\rm  
em})}\right\},\eqno(5b)$$
where $r_{\rm em}$ is the radius at which the neutrino is produced  
(the
neutrino sphere). The redshift at Schwarzschild radial coordinate $r$  
is then
$z_r = {E_{\rm em}}/{E^{\ast} (r)} -1$, so that $1+z_r =  
{\exp}{\left\{{\Phi
(r) -\Phi (r_{\rm em})}\right\}}$, and the redshift at radial  
infinity is $z
\equiv z_{\infty} = {\left({1 - r_s/r_{\rm em}}\right)}^{-1/2} -1$.

At Schwarzschild radial coordinate $r$, the electron fraction is  
determined as
in equations $(2a)$
and $(2b)$ by the {\it local} values of the luminosities and average  
energies
of the ${\bar{\nu}}_e$ and ${\nu}_e$. However, since these neutrino  
species
have differing production/emission radii ({\it i.e.}, their neutrino  
spheres
have different values of Schwarzschild radial coordinate), they are  
subject to
different gravitational redshift effects. If we define the ${\nu}_e$  
neutrino
sphere to be at $r^{\rm sp}_{{\nu}_e}$ and the ${\bar{\nu}}_e$  
neutrino sphere
to be at $r^{\rm sp}_{{\bar{\nu}}_e}$, then equation $(2b)$ can be  
recast as:
$$Y_e = 1/\left\{1 + R_{n/p} \right\};\eqno(6a)$$
$$R_{n/p} \equiv R^0_{n/p} \cdot {\Gamma};\eqno(6b)$$
$$R^0_{n/p} \approx \left[{L^{\rm sp}_{\bar\nu_e}  
\langle E^{\rm
sp}_{\bar\nu_e}\rangle\over
L^{\rm sp}_{\nu_e} \langle E^{\rm sp}_{\nu_e}\rangle
}\right].\eqno(6c)$$
In these equations, $\langle E^{\rm sp}_{{\bar{\nu}}_e}\rangle $ and  
$L^{\rm
sp}_{{\bar{\nu}}_e}$ are the average ${\bar{\nu}}_e$ energy and  
luminosity as
measured by a locally inertial observer at rest at the  
${\bar{\nu}}_e$ neutrino
sphere; and similarly for the quantities which characterize the  
${\nu}_e$
energy and luminosity at the ${\nu}_e$ neutrino sphere. We are making  
the
approximation that the (anti)neutrino energy spectrum does not evolve
significantly with increasing radius above the (anti)neutrino sphere  
as a
result of emission, absorption and scattering processes. The quantity  
$R_{n/p}$
is the local neutron-to-proton ratio, and as is evident from equation  
$(2a)$,
this can be approximated as
$R_{n/p} \approx {\lambda}_{\bar\nu_e p}(r)/{\lambda}_{\nu_e  
n}(r)$,
where the rates are understood to be evaluated from the neutrino and
antineutrino energy spectra extant at Schwarzschild radial coordinate  
$r$.  In
equation $(6b)$, $R^0_{n/p}$ can be interpreted as the  
neutron-to-proton ratio
[corresponding to the electron fraction $Y^0_e = 1/(1+R^0_{n/p})$] in  
the
absence of gravitational redshift effects. Here the effects of the
gravitational field are represented by the term ${\Gamma}$ which can  
be written
as,
$$\Gamma \equiv {\exp}\left\{3\left[\Phi\left(r^{\rm
sp}_{\bar\nu_e}\right) - \Phi\left(r^{\rm
sp}_{\nu_e}\right)\right]\right\},\eqno(7a)$$
$$\Gamma =\left({1-r_s/r^{\rm
sp}_{\bar\nu_e}\over 1-r_s/r^{\rm sp}_{\nu_e}
}\right)^{3/2},\eqno(7b)$$
where equation $(7b)$ obtains in vacuum.
Note that ${\Gamma}$ contains three factors of the redshift term  
which appears
in equations $(5a)$ and $(5b)$. This is because the product of  
neutrino
luminosity and average energy has contained in it two \lq\lq  
energy\rq\rq\
factors and one inverse time interval ({\it i.e.}, frequency), which  
redshifts
in the same manner as energy.

Let us define the difference of the neutrino and antineutrino sphere  
positions
to be ${\delta}r \equiv r^{\rm sp}_{{\nu}_e} - r^{\rm  
sp}_{{\bar{\nu}}_e}$. If
we now assume that ${\delta}r << r^{\rm sp}_{{\bar{\nu}}_e}$, then we  
can
estimate that,
$$R_{n/p} \approx R^0_{n/p} \cdot {\left( 1 +
{{3}\over{2}}{\gamma} {{{\delta}r}\over{r^{\rm sp}_{{\bar{\nu}}_e}}}
\right)}^{-1},\eqno(8a)$$
where the redshift amplification factor is defined to be,
$$\gamma \equiv 1/{\left( {r^{\rm sp}_{{\bar{\nu}}_e}}/{r_s}
-1\right)}.\eqno(8b)$$
For example, if ${\delta}r/r_{\bar\nu_e}^{\rm sp} \approx 0.1$ and  
the
predicted electron fraction in the absence of gravitational redshift  
effects is
$Y^0_e \approx 0.4$, then we must have ${r_s}/{r^{\rm  
sp}_{{\bar{\nu}}_e}} <
0.76$ if we are to have more neutrons than protons, $Y_e < 0.5$ when
gravitational redshift effects are taken into account. In fact, the
differential redshift effect is even greater when the threshold  
effects are
taken into account in the evaluation of the rates which enter into  
$R_{n/p}
\approx {\lambda}_{\bar\nu_e p}(r)/{\lambda}_{\nu_e n}(r)$.  
Since the
neutron is heavier than the proton by $\Delta_{np}\approx 1.293\,{\rm  
MeV}$,
the process in equation $(1b)$ has a threshold, and any reduction in  
the
average energies of the neutrinos and antineutrinos has a  
disproportionate
effect on reducing ${\lambda}_{{\bar{\nu}}_e p}(r)$. A semi-numerical
calculation including these threshold effects shows that for  $L^{\rm
sp}_{{\bar{\nu}}_e}/L^{\rm sp}_{{\nu}_e} \approx 1.2$, and
${\epsilon_{\bar\nu_e}^{\rm sp}\equiv\langle{(E^{\rm
sp}_{{\bar{\nu}}_e})^2}\rangle}/\langle E^{\rm sp}_{\bar\nu_e}\rangle  
\approx
25\,{\rm MeV}$ and ${\epsilon_{\nu_e}^{\rm sp}\equiv\langle{(E^{\rm
sp}_{{\nu}_e}})^2\rangle}/\langle E_{\nu_e}^{\rm sp}\rangle \approx  
10\,{\rm
MeV}$ (which imply that $Y^0_e \approx 0.32$), and ${\delta}r/r^{\rm
sp}_{{\bar{\nu}}_e} \approx 0.1$, we must have ${r_s}/r^{\rm
sp}_{{\bar{\nu}}_e} < 0.58$ if we are to guarantee that the electron  
fraction
is $Y_e < 0.4$. The same parameters, but now assuming that  
${\delta}r/r^{\rm
sp}_{{\bar{\nu}}_e} \approx 0.01$, would require ${r_s}/r^{\rm
sp}_{{\bar{\nu}}_e} < 0.75$ if we demand that $Y_e < 0.4$. A value of  
$Y_e
\approx 0.4$ may actually be a conservative limit on the maximum  
allowable
electron fraction for the $r$-process.

We see that {\it plausible} values of the neutron star mass (or  
$r_s$), radius
($\approx r^{\rm sp}_{{\bar{\nu}}_e}$) and density scale height near  
the
surface (which determines $\delta r$) could produce enough  
differential
gravitational redshift to {\it preclude} neutrino-heated supernova  
ejecta as
the site of the $r$-process. If eventually it is argued convincingly  
that this
site {\it does} give rise to the $r$-process, then differential  
neutrino
redshift could allow the nucleosynthesis abundance yield to become an
interesting probe of the nuclear equation of state parameters that  
set $r^{\rm
sp}_{{\bar{\nu}}_e}$ and $\delta r$. Clearly, the differential  
redshift effects
become more dramatic as $\gamma$ becomes larger and the star becomes  
more
relativistic. In turn, we know from the work of Gerry Brown and  
others that a
phase transition to a soft kaon condensed environment (or interacting  
strange
quark matter) will cause the star to become more relativistic. One  
might argue
that, for a particularly soft equation of state, $\delta r$ might  
approach
zero, and the neutrinos and antineutrinos could then decouple from  
the same
radius with (possibly) the same energy spectrum, thus ensuring that  
there will
be no differential redshift effect. However, this particular case  
will also
preclude $r$-process nucleosynthesis, as nearly identical energy  
spectra for
the ${\bar{\nu}}_e$ and the ${\nu}_e$ will produce $Y_e > 0.5$ as a  
result of
the energy threshold for the process in equation $(1b)$. In any case,  
the
effects of an exotic state of matter on the neutrino and antineutrino  
energy
spectra call for further examination.

This work was supported by NSF Grant PHY95-03384 and NASA Grant  
NAG5-3062 at
UCSD. Y.-Z. Qian was supported by the D. W. Morrisroe Fellowship at  
Caltech.

\end{document}